\documentclass[twocolumn,showpacs,preprintnumbers,amsmath,amssymb]{revtex4-1}

\usepackage{graphicx} \usepackage{amssymb} \usepackage{hyperref} \textwidth 17.5cm
\begin{document}

\title{Suppression of Shastry-Sutherland phase driven by electronic concentration reduction  in magnetically frustrated Ce$_2$Pd$_2$Sn$_{1-y}$In$_y$ alloys}

\author{J.G. Sereni$^1$, J. Roberts$^2$, F. Gastaldo$^2$, M. Giovannini$^2$}
\address{$^1$Low Temperature Division, CAB-CNEA and CONICET, 8400 San
Carlos de Bariloche, Argentina\\
$^2$Dipartimento di Chimica, Universita' di Genova, Via Dodecaneso 33, I-16146 Genova, Italy}

\begin{abstract}

{Shastry-Sutherland lattice was observed as alternative ground state in Rare Earth intermetallic with Mo$_2$B$_2$Fe and U$_2$Pt$_2$Sn anisotropic structures where magnetic frustration is favored. In the case 
of Ce$_2$Pd$_2$Sn, it was shown that such phase can be suppressed by the application of magnetic field  and, in this work, its stability is studied as a function of the electronic concentration by doping the 
Sn(4+) lattice with In(3+) atoms. Magnetic and specific heat measurements show that around 50\% substitution the Shastry Sutherland lattice vanishes in a critical point. This result confirms the strong dependence of that 
phase on the electron density because a recent investigation on the Pd rich solid solution Ce$_{2+\epsilon}$Pd$_{2-\epsilon}$In$_{1-x}$Sn$_x$ (with $\epsilon < 0$) demonstrates that atomic disorder dominates 
the phase diagram at intermediate Sn/In concentration inhibiting magnetic frustration effects. In the alloys investigated in this work, the $\epsilon >0$ character stabilizes  the ferromagnetic ground state all along the concentration, allowing 
the Shastry Sutherland lattice formation on the Sn rich side.}

\end{abstract} \date{\today} \maketitle

 \section{Introduction}

Intermetallic compounds with the formula $R_2T_2X$ (with $R  =$ rare earths, $T  =$ transition  metals and $X =$ Sn, In, Pb and Cd)  have received considerable attention due to their varied behaviors, such as 
intermediate valence \cite{a}, non-Fermi liquid \cite{b,Yb2Sn/In} and quantum phase transitions \cite{Yb2Sn/In,d}. Most of these compounds crystallize in the tetragonal Mo$_2$FeB$_2$-type \cite{Fourgeot}, whereas a few 
of them are formed in a related variant U$_2$Pt$_2$Sn-type. Both structures consist of alternate $R$-planes and $T /X$–planes stacked along the $c$ axis. In the first layers, magnetic $R$ atoms form a 
network of triangles and squares (mimicking a pinwheel mosaic). This geometry provides the ideal conditions for magnetic frustration and the consequent formation of the Shastry-Sutherland phase as 
alternative ground state, provided that the interactions between next neighbors may form a square lattice of magnetic dimers. In fact, Shastry-Sutherland (ShSu) lattices \cite{ShSu} were observed in intermetallic 
compounds like Yb$_2$Pt$_2$Pb \cite{Aronson, Aronson2} crystallizing in the U$_2$Pt$_2$Sn-type and Ce$_2$Pd$_2$Sn  \cite{Ce2Pd2Sn} with the Mo$_2$FeB$_2$-type.

In the case of Ce$_2$Pd$_2$Sn the ShSu phase  is detected between a temperature ($T_S$) at which neighboring Ce atoms form magnetic dimers distributed at the corners of a simple square 2D lattice and a FM transition 
($T_C$) at which inter-plane interactions transform the system into a 3D magnet  \cite{Ce2Pd2Sn}. The magnetic stability of this phase in Ce$_2$Pd$_2$Sn was previously studied under magnetic field \cite{supress}, detecting that it is suppressed at a critical field $B_{cr}=0.11\pm0.01$\,T. 

In order to recognize the stability of that phase as a function of electronic concentration, the progressive substitution of tetravalent (4+) Sn atoms by trivalent (3+) In ones was investigated. The isotypic  Ce$_2$Pd$_2$In is known to form in the same  Mo$_2$B$_2$Fe-type structure \cite{Fourgeot}. However, instead of a stoichiometric compound it was determined that it forms as a solid solution Ce$_{2+\epsilon}$Pd$_{2-
\epsilon}$In \cite{2000} around that concentration. Moreover, the low temperature magnetic properties were recognized to depend on the relative excess of Ce ($\epsilon > 0$)  or Pd ($\epsilon < 0$) resulting in  ferromagnetic (FM) or antiferromagnetic (AFM) ground states (GS) respectively \cite{2000,ElectrConc}. These two branches on the $In$ rich side allows to study the ShSu lattice stability within the Ce$_2$Pd$_2$Sn$_{1-y}
$In$_y$ alloys following to comparable trails, one pointing to a FM and the other to an AFM final GS. 

Recently, the $\epsilon < 0$ branch, i.e. starting from the AFM sate in Ce$_2$Pd$_2$(In$_{1-x}$Sn$_x$), was reported  \cite{InAFM} to show relevant atomic disorder for $x\geq 0.4$ that inhibits the formation of 
the ShSu lattice at higher Sn content. Such magnetic disorder is reflected in a broadening of the $T_N(x)$ transition at that concentration.  
Complementary, in this work we report on the stability of the ShSu phase following the rich Ce, Ce$_{2.15}$Pd$_{1.95}$Sn$_{1-y}$In$_y$ FM branch, i. e. with $\epsilon > 0$, that forms continuously along the full Sn/In composition and show well defined magnetic transitions.

\section{Experimental details and results} 
\subsection{Sample Preparation and Characterization}

The samples were prepared by weighing the proper amounts of elements, then arc melted in an Argon atmosphere on a water cooled copper hearth with a tungsten electrode. To ensure good
homogeneity the buttons were turned over and remelted several times. Weight losses after melting were always smaller than 0.5 mass percent. Later, the samples were annealed at 750\,$^o$C for
ten days and quenched in cold water. Scanning electron microscopy (SEM), and electron probe micro-analysis (EPMA) based on energy-dispersive X-ray spectroscopy were used to examine phase
compositions. The compositional contrast was revealed in unetched samples by means of a backscattered electron detector (BSE). The composition values derived were usually accurate to 1\,at\%. X-ray
diffraction (XRD) was performed on powder samples using a vertical diffractometer X-Pert with Cu\,K$_{\alpha}$ radiation. Lattice parameters slightly change along the concentration range from $a=7.765\AA$ and $c=3.902 \AA$, with a $c/a$ =0.5026 ratio of the $x = 0$ mother compound \cite{Ce2Pd2Sn}. 

DC-magnetization measurements were carried out using a standard SQUID magnetometer operating between 1.8 and 320\,K, and as a function of field up to 5\,T. Specific heat was measured using a standard 
heat pulse technique in a semi-adiabatic He-3 calorimeter in the range between 0.5 and
20\,K, at zero applied magnetic field.

\begin{figure}[tb] \begin{center} \includegraphics[width=20pc]{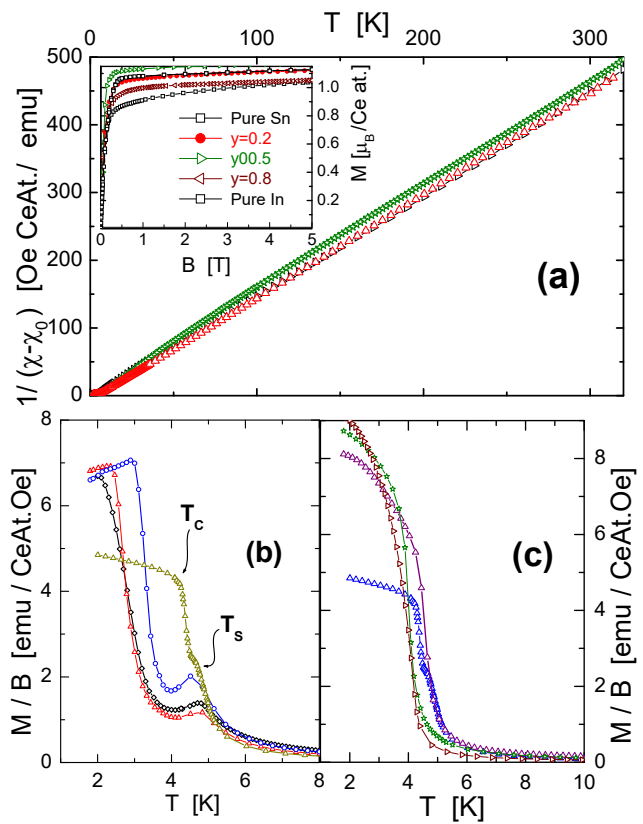}
\end{center} \caption{(a) Inverse magnetic susceptibility from 1.8\,K up to room temperature measured in a field $B=1$\,T. Inset: magnetization versus field up to 5\,T, measured at 1.8\,K. (b) Low temperature susceptibility of Sn rich alloys, measured with $B=100$\,Oe, showing how the ShSu phase is progressively suppressed by increasing In content. (c) The same for the In rich concentrations.}
\label{F1} \end{figure}

\subsection{Magnetic Properties}

High temperature ($T>30$\,K) magnetic susceptibility results are
properly described by a $\chi=\chi_{cw}+\chi_p$ dependence, see Fig.~\ref{F1}a, where
the first term corresponds to the temperature dependent Curie-Weiss contribution $\chi_{cw}= \frac{Cc}{T+\theta}$ and the second to a Pauli-like contribution. This weak $\chi_p$ contribution is
observed along the full concentration range with values ranging within $\chi_p=0.6\pm 0.4) 10^{-3}$\,emu/Ce\,at.Oe. From the inverse of $\chi_{cw}$ one may extract the Curie constant
(Cc) which indicates a large but not fully developed Ce$^{3+}$ magnetic
moment (i.e. $\mu_{eff} = 2.32\,\mu_B$ per Ce atom). The paramagnetic temperature extrapolated from high temperature ($T>50$\,K) is notably weak $\theta_P \leq 7$\,K all along the concentration variation, 
excluding any relevant effect of local moments-conduction band hybridization or Kondo effect.

On the stoichiometric $Sn$ limit, the six fold degenerated state established by Hund's rules for the $J=5/2$
angular momentum of Ce atoms is split by the crystal electric field (CEF) effect into three Kamer's doublets 
with the excited ones at $\Delta_1 \approx 65$\,K and $\Delta_1 \approx 230$\,K \cite{Ce2Pd2Sn}. Similar 
value ($\Delta_1 \approx 60$\,K) was reported for the first excited doublet on the In side 
\cite{ElectrConc}. This coincidence indicates that no relevant modification of the CEF occurs along the Sn/In 
substitution guaranteeing that the low temperature magnetic properties is governed by a pure doublet GS. 

Magnetization $M(B)$ curves at $T=1.8$\,K are shown in the inset of 
Fig.~\ref{F1}a. The saturation values slightly increases from $1.02\,\mu_B$ to $1.13\,\mu_B$ between pure $Sn$ and $In$ limits, with a maximum value at the critical concentration $y_{cr} = 0.5$  of  $1.18\,\mu_B$.

The low temperature magnetization normalized by field ($M/B$) is presented in Fig.~\ref{F1}b including
the $0\leq y \leq 0.5$ samples where both  $T_S(y)$ and $T_C(y)$ transitions are clearly identified. Fig.~\ref{F1}c contains the FM magnetization of the $0.5 \leq y \leq 1$ samples. 

 \begin{figure}[tb] \begin{center} \includegraphics[width=19pc]{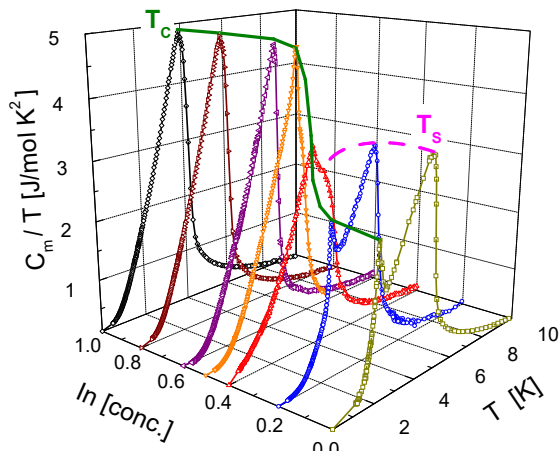}
\end{center} \caption{Magnetic contribution to the specific heat of the measured alloys in a 3D representation. T$_C(y)$ indicates the FM transition (continuous curve) and $T_S(y < 5)$ the transition into the Shastry Sutherland phase (dashed curve)}
\label{F2} \end{figure}

\subsection{Specific Heat}

To extract the magnetic contribution ($C_m$) to the measured specific heat ($C_P$), the La$_{2}$Pd$_2$In compound was used as a reference for the phonon subtraction \cite{Ce2Pd2Sn} as $C_m = C_P(T) 
- C(La_2Pd_2In)$. On the $Sn$-rich side, the two magnetic transitions observed in magnetic 
measurements (see Fig.~\ref{F1}b) are reflected in corresponding $C_m(T)/T$ jumps presented in 
Fig.~\ref{F2}. On the $Sn$-rich side, the specific heat jump at the $T= T_S$  (dashed curve)  transition 
between paramagnetic and ShSu phase exhibits a second order character, whereas the lower one ($T_C$) 
associated to the ferromagnetic phase shows the characteristic of a first order.  However, above the critical 
concentration ($y_{cr} \approx 0.5$), at which the ShSu phase is suppressed, the only transition $T_C$ 
becomes of second order keeping the tail originated in magnetic fluctuations arising above the transition. 

Coincidentally with the previous study on stoichiometric Ce$_2$Pd$_2$Sn, the ShSu phase suppression by magnetic field,  $T_S(y)$ slightly changes in temperature with $In$ concentration $T_S(0\leq y \leq 0.5) = (4.5\pm 0.2$\,K) whereas $T_C(y=0) = 2.1$\,K increases from up to reach the critical point at $T_{cr}(y_{cr}) = 4.15$\,K.  

Notably, the height  of $C_m/T$ at $T_C(0\leq y < 0.5)$ is clearly larger than for $T_S(0.5 \leq y \leq 1)$. 
This indicates that the transition at $T=T_S$ corresponds to an intermediate phase that involves a partial condensation of the available degrees of freedom which are then condensed into the FM phase at $T< T_C$. On the contrary,  the transition at for $x \geq y_{cr}$ involves all magnetic degrees of freedom.

\begin{figure}[tb] \begin{center} \includegraphics[width=19pc]{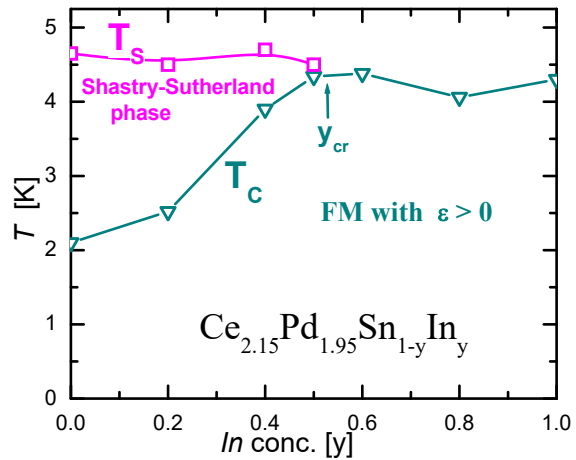}
\end{center} \caption{Magnetic phase diagram showing the evolution of the  Curie ($T_C$) and Shastry-Sutherland ($T_S$) transitions as a function of In increase.}
\label{F3} \end{figure}

\section{Discussion}

\begin{figure}[tb] \begin{center} \includegraphics[width=18pc]{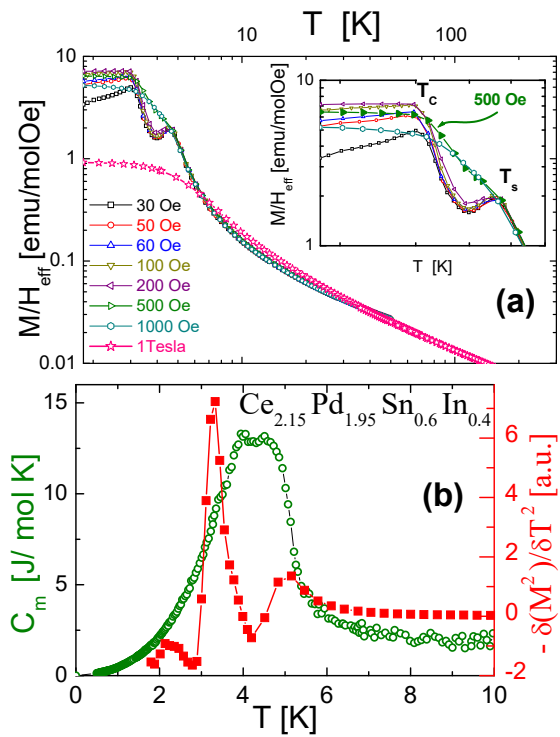}
\end{center} \caption{(a) Progressive Shastry Sutherland phase suppression by magnetic field at 40\% In in a zero field cooling process. Inset: detail of the mentioned suppression. (b) Comparison between specific heat measurement and magnetization derivative.}
\label{F4} \end{figure}

\subsection{Magnetic phase diagram}

Based on magnetic and specific heat results, the evolution of the Shastry-Sutherland ($T_S$) and Curie ($T_C$) transitions can be traced as a function of $In$ increase (upper FM curves). One can see that the Shastry-Sutherland holds up to about 50\% of$In$ substitution.

On the $Sn$-rich side, the characteristics of the ShSu phase recognized in the Ce$_2$Pd$_2$Sn compound \cite{Ce2Pd2Sn} can be traced up to the sample with 40\% of In 
concentration. Although the difference between $T_S$ and $T_C$ in the specific heat measurements is observed as a structure in a unique anomaly, the $M(T)$ dependence allows to better distinguish both 
transitions at that concentration, see  Fig.\ref{F4}a. Both properties are better compared in the lower panel (Fig.\ref{F4}b), using as magnetic parameter the second derivative because in FM systems the internal magnetic energy $U_m \propto M^2$ and therefore $C_m \propto \delta M^2/\delta ^2T$.

Another identification of the ShSu phase can be done through its suppression by a  
magnetic field, which occurs at $B_{cr}(y=0.4) \approx 500$\,Oe, see the inset in Fig.\ref{F4}a. This behavior can be compared with the suppression of the ShSu phase of the stoichiometric compounds Ce
$_2$Pd$_2$Sn  \cite{Ce2Pd2SnField}, taking into account that $B_{cr}(y=0.4) \approx 1/2 B_{cr}(y=0)$.

\subsection{Critical point} 

\begin{figure}[tb] \begin{center} \includegraphics[width=20pc]{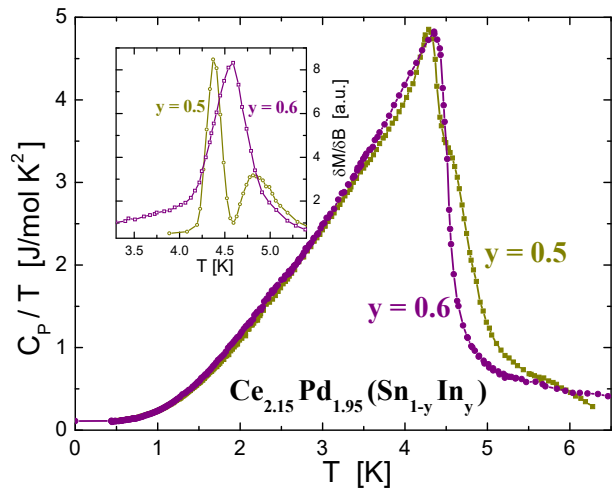}
\end{center} \caption{Critical concentration determined by specific heat and, in the inset, by the magnetization derivatives of samples  $y = 0.5$ and $y = 0.6$. }
\label{F5} \end{figure}

The full suppression of the ShSu phase by increasing $In$ concentration is observed slightly above the 50\% of substitution, where the $T_S(y)$ transition joints $T_C(y)$ FM phase boundary. In Fig.\ref{F5} the 
specific heat transitions of two neighboring concentrations, $y = 0.5$ and $y = 0.6$, are compared making evident that the later does not show any structure around its maximum which is still detected in the $x = 0.5$  
one. Coincidentally, there is practically no difference between respective entropies collected up to the transition.

In the inset of Fig.\ref{F5} the same comparison is performed on respective magnetization derivatives making the difference more evident. Notably, the first order type cusp in the $y =0.5$ sample fully disappears at the following concentration $y=0.6$. 
 
\subsection{Physical behavior beyond the critical point} 

\begin{figure}[tb] \begin{center} \includegraphics[width=20pc]{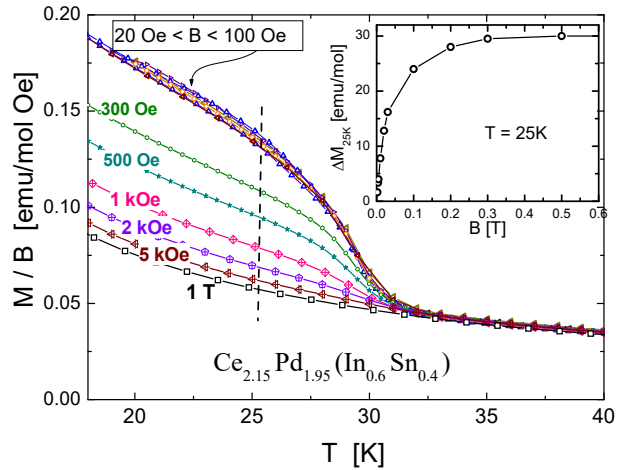}
\end{center} \caption{Field dependence of the unexpected ferromagnetic contribution below about 30\,K tentatively attributed to spin density waves formation. Inset: Magnetization increase upon the bulk magnetization at 25K.}
\label{F6} \end{figure}

Between $y=0.6$ and 1, the typical FM $M(T)$ dependence is observed, see  Fig.\ref{F1}c, while $C_m/T$ shows the same behavior like for pure $In$. Independently of these low temperature features, sample $x = 
0.8$ shows an unexpected ferromagnetic signal in the thermal dependence of its magnetic susceptibility 
arising at $T\leq 30$\,K, see  Fig.\ref{F6}.
This FM contribution was analyzed as a function of magnetic field at $T=25$\,K reflected in an increase of the measured magnetization with field $\Delta M_{25\,K}(B)$,
see the inset in Fig.\ref{F6}. This FM signal is found to saturate at around 0.3\,T. 
Such contribution cannot be explained by any foreign phase constituted by any combination of 
the four elements Ce-Pd-Sn-In, therefore it could be more likely related to the onset of spin density waves. A microscopic investigation on the nature of this anomaly is required to confirm its origin.\

\section{Conclusions}

This study demonstrates that, contrary to the AFM ( $\epsilon < 0$) branch of Ce$_{2+\epsilon} $Pd$_{2- \epsilon}$(In$_{1-x}$Sn$_x$)  alloys, the FM ($\epsilon > 0$) one shows well defined 
transitions with irrelevant atomic disorder produced by Sn/In substitution. As a consequence the upper ($T=T_S$) and the lower  ($T=T_C$) boundaries of the ShSu phase can be clearly traced up to $y \approx 50\%$ of Sn/In substitution. The reduction of the electronic concentration driven by doping the Sn(4+) lattice with In(3+) atoms progressively weakens the stability of that phase reflected in the monotonous growing of the FM phase in detriment of the former and the consequent increase of $T_C(y)$. 
Magnetic field effect, analyzed at the edge of the critical concentration $y=0.4$, shows that only 1/2 of field is required to suppress the ShSu phase in comparison to stoichiometric Ce$_2$Pd$_2$Sn. 

Although Ce rich content ($\epsilon >0$) reduces atomic disorder effects allowing to extend the ShSu phase range, further decrease of electronic concentration after Sn/In substitution definitively suppresses the possibility of dimers formation on which simple square ShSu lattice builds up. Since this lattice forms within the Ce planes, one may expect that on the $In$-rich side (i.e. $y > 0.5$) interplane interactions arise making the system to behave like a 3D - FM as soon as the magnetic order parameter arises.

\end{document}